\documentclass[manyauthors,nocleardouble,COMPASS]{cernphprep}

\usepackage[utf8]{inputenc}
\usepackage{amsmath}
\usepackage{bm}		
\usepackage{cite}
\usepackage{appendix}
\usepackage{lineno}
\usepackage{bigstrut}
\usepackage{ulem}

\newcommand\varpm{\mathbin{\vcenter{\hbox{%
  \oalign{\hfil$\scriptstyle+$\hfil\cr
          \noalign{\kern-.3ex}
          $\scriptscriptstyle\{-\}$\cr}%
}}}}

\pdfoutput=1    

\usepackage[pdftex]{hyperref}
\hypersetup{
  pdftitle={Probing transversity by measuring $\Lambda$ polarisation in SIDIS},%
  pdfauthor={The COMPASS Collaboration},%
  pdfsubject={},%
  pdfkeywords={},%
  pdfstartview={},%
  bookmarksopen=true, breaklinks=true, debug=true, %
  colorlinks=true, linkcolor=blue, citecolor=blue, urlcolor=blue
}
\begin{document}
\begin{titlepage}
\PHnumber{2020--ver.5.0}
\PHdate{\today}
\DEFCOL{CDS-Library}

\title{Probing transversity by measuring $\Lambda$ polarisation in SIDIS}
\Collaboration{The COMPASS Collaboration}
\ShortAuthor{The COMPASS Collaboration}

\begin{abstract}
Based on the observation of sizeable target-transverse-spin asymmetries in single-hadron and hadron-pair production in Semi-Inclusive measurements of Deep Inelastic Scattering (SIDIS), the chiral-odd transversity quark distribution functions 
$h_1^q$ are nowadays well established.
Several possible channels to access these functions were
originally proposed. One candidate is the measurement of the polarisation of $\Lambda$ hyperons produced in SIDIS off transversely polarised nucleons, where the transverse polarisation of the struck quark might be transferred to the final-state hyperon.
In this article, we present the COMPASS results on the  transversity-induced polarisation of $\Lambda$ and $\bar{\Lambda}$ hyperons produced in SIDIS off transversely polarised protons. Within the experimental uncertainties, no significant deviation from zero was observed. 
The results are discussed in the context of different models taking into account previous experimental results on 
$h_1^u$ and $h_1^d$.
\end{abstract}
\Submitted{(to be submitted to Physics Letters B)}

\end{titlepage}

{
\pagestyle{empty}
%
%
\section*{The COMPASS Collaboration}
\label{app:collab}
\renewcommand\labelenumi{\textsuperscript{\theenumi}~}
\renewcommand\theenumi{\arabic{enumi}}
\begin{flushleft}
G.D.~Alexeev\Irefn{dubna},
M.G.~Alexeev\Irefnn{turin_u}{turin_i},
A.~Amoroso\Irefnn{turin_u}{turin_i},
V.~Andrieux\Irefnn{cern}{illinois},
V.~Anosov\Irefn{dubna},
K.~Augsten\Irefnn{dubna}{praguectu},
W.~Augustyniak\Irefn{warsaw},
C.D.R.~Azevedo\Irefn{aveiro},
B.~Bade{\l}ek\Irefn{warsawu},
F.~Balestra\Irefnn{turin_u}{turin_i},
M.~Ball\Irefn{bonniskp},
J.~Barth\Irefn{bonniskp},
R.~Beck\Irefn{bonniskp},
Y.~Bedfer\Irefn{saclay},
J.~Berenguer~Antequera\Irefnn{turin_u}{turin_i},
J.~Bernhard\Irefnn{mainz}{cern},
M.~Bodlak\Irefn{praguecu},
F.~Bradamante\Irefn{triest_i},
A.~Bressan\Irefnn{triest_u}{triest_i},
V.E.~Burtsev\Irefn{tomsk},
W.-C.~Chang\Irefn{taipei},
C.~Chatterjee\Irefnn{triest_u}{triest_i},
M.~Chiosso\Irefnn{turin_u}{turin_i},
A.G.~Chumakov\Irefn{tomsk},
S.-U.~Chung\Irefn{munichtu}\Aref{B}\Aref{B1},
A.~Cicuttin\Irefn{triest_i}\Aref{C},
P.M.M.~Correia\Irefn{aveiro},
M.L.~Crespo\Irefn{triest_i}\Aref{C}
% ...
\end{flushleft}
%
%
\begin{Authlist}
\item \Idef{aveiro}{University of Aveiro, I3N, Dept.\ of Physics, 3810-193 Aveiro, Portugal}
\item \Idef{bochum}{Universit\"at Bochum, Institut f\"ur Experimentalphysik, 44780 Bochum, Germany\Arefs{Q}\Aref{R}}
\item \Idef{bonniskp}{Universit\"at Bonn, Helmholtz-Institut f\"ur  Strahlen- und Kernphysik, 53115 Bonn, Germany\Arefs{Q}}
\item \Idef{bonnpi}{Universit\"at Bonn, Physikalisches Institut, 53115 Bonn, Germany\Arefs{Q}}
\item \Idef{brno}{Institute of Scientific Instruments of the CAS, 61264 Brno, Czech Republic\Arefs{S}}
\item \Idef{calcutta}{Matrivani Institute of Experimental Research \& Education, Calcutta-700 030, India\Arefs{T}}
\item \Idef{dubna}{Joint Institute for Nuclear Research, 141980 Dubna, Moscow region, Russia\Arefs{T1}}
\item \Idef{freiburg}{Universit\"at Freiburg, Physikalisches Institut, 79104 Freiburg, Germany\Arefs{Q}\Aref{R}}
\item \Idef{cern}{CERN, 1211 Geneva 23, Switzerland}
\item \Idef{liberec}{Technical University in Liberec, 46117 Liberec, Czech Republic\Arefs{S}}
\item \Idef{lisbon}{LIP, 1649-003 Lisbon, Portugal\Arefs{U}}
\item \Idef{mainz}{Universit\"at Mainz, Institut f\"ur Kernphysik, 55099 Mainz, Germany\Arefs{Q}}
\item \Idef{miyazaki}{University of Miyazaki, Miyazaki 889-2192, Japan\Arefs{V}}
\item \Idef{moscowlpi}{Lebedev Physical Institute, 119991 Moscow, Russia}
\item \Idef{munichtu}{Technische Universit\"at M\"unchen, Physik Dept., 85748 Garching, Germany\Arefs{Q}\Aref{D}}
\item \Idef{nagoya}{Nagoya University, 464 Nagoya, Japan\Arefs{V}}
\item \Idef{praguecu}{Charles University, Faculty of Mathematics and Physics, 18000 Prague, Czech Republic\Arefs{S}}
\item \Idef{praguectu}{Czech Technical University in Prague, 16636 Prague, Czech Republic\Arefs{S}}
\item \Idef{protvino}{State Scientific Center Institute for High Energy Physics of National Research Center `Kurchatov Institute', 142281 Protvino, Russia}
\item \Idef{saclay}{IRFU, CEA, Universit\'e Paris-Saclay, 91191 Gif-sur-Yvette, France\Arefs{R}}
\item \Idef{taipei}{Academia Sinica, Institute of Physics, Taipei 11529, Taiwan\Arefs{W}}
\item \Idef{telaviv}{Tel Aviv University, School of Physics and Astronomy, 69978 Tel Aviv, Israel\Arefs{X}}
\item \Idef{tomsk}{Tomsk Polytechnic University, 634050 Tomsk, Russia\Arefs{Y}}
\item \Idef{triest_u}{University of Trieste, Dept.\ of Physics, 34127 Trieste, Italy}
\item \Idef{triest_i}{Trieste Section of INFN, 34127 Trieste, Italy}
\item \Idef{turin_u}{University of Turin, Dept.\ of Physics, 10125 Turin, Italy}
\item \Idef{turin_i}{Torino Section of INFN, 10125 Turin, Italy}
\item \Idef{illinois}{University of Illinois at Urbana-Champaign, Dept.\ of Physics, Urbana, IL 61801-3080, USA\Arefs{Z}}
\item \Idef{warsaw}{National Centre for Nuclear Research, 02-093 Warsaw, Poland\Arefs{a} }
\item \Idef{warsawu}{University of Warsaw, Faculty of Physics, 02-093 Warsaw, Poland\Arefs{a} }
\item \Idef{warsawtu}{Warsaw University of Technology, Institute of Radioelectronics, 00-665 Warsaw, Poland\Arefs{a} }
\item \Idef{yamagata}{Yamagata University, Yamagata 992-8510, Japan\Arefs{V} }
\end{Authlist}
%
%
\renewcommand\theenumi{\alph{enumi}}
\begin{Authlist}
\item [{\makebox[2mm][l]{\textsuperscript{\#}}}] 
Corresponding authors\\
{\it E-mail addresses}: Oleg.Denisov@cern.ch, Fulvio.Tessarotto@cern.ch\\
\item \Adef{B}{Also at Dept.\ of Physics, Pusan National University, Busan 609-735, Republic of Korea}
\item \Adef{B1}{Also at Physics Dept., Brookhaven National Laboratory, Upton, NY 11973, USA}
\item \Adef{C}{Also at Abdus Salam ICTP, 34151 Trieste, Italy}
\item \Adef{D}{Supported by the DFG cluster of excellence `Origin and Structure of the Universe' (www.universe-cluster.de) (Germany)}
\item \Adef{F}{Supported by the DFG Research Training Group Programmes 1102 and 2044 (Germany)}
\item \Adef{G}{Also at Chubu University, Kasugai, Aichi 487-8501, Japan}
\item \Adef{H}{Also at Dept.\ of Physics, National Central University, 300 Jhongda Road, Jhongli 32001, Taiwan}
\item \Adef{I}{Also at KEK, 1-1 Oho, Tsukuba, Ibaraki 305-0801, Japan}
\item \Adef{K}{Also at Moscow Institute of Physics and Technology, Moscow Region, 141700, Russia}
\item \Adef{K1}{Deceased}
\item \Adef{L}{Also at Yerevan Physics Institute, Alikhanian Br. Street, Yerevan, Armenia, 0036}
\item \Adef{M}{Also at Dept.\ of Physics, National Kaohsiung Normal University, Kaohsiung County 824, Taiwan}
\item \Adef{M1}{Supported by ANR, France with P2IO LabEx (ANR-10-LBX-0038) in the framework ``Investissements d'Avenir'' (ANR-11-IDEX-003-01)}
\item \Adef{N}{Also at Institut f\"ur Theoretische Physik, Universit\"at T\"ubingen, 72076 T\"ubingen, Germany}
\item \Adef{N1}{Retired}
\item \Adef{N2}{Present address: Brookhaven National Laboratory, Brookhaven, USA}
\item \Adef{O}{Also at University of Eastern Piedmont, 15100 Alessandria, Italy}
\item \Adef{P}{Present address: RWTH Aachen University, III.\ Physikalisches Institut, 52056 Aachen, Germany}
\item \Adef{P1}{Present address: Universit\"at Hamburg, 20146 Hamburg, Germany}
\item \Adef{Q}{Supported by BMBF - Bundesministerium f\"ur Bildung und Forschung (Germany)}
\item \Adef{R}{Supported by FP7, HadronPhysics3, Grant 283286 (European Union)}
\item \Adef{S}{Supported by MEYS, Grant LM20150581 (Czech Republic)}
\item \Adef{T}{Supported by B.~Sen fund (India)}
\item \Adef{T1}{Supported by CERN-RFBR Grant 12-02-91500}
\item \Adef{U}{Supported by FCT, Grants CERN/FIS-PAR/0007/2017 and  CERN/FIS-PAR/0022/2019 (Portugal)}
\item \Adef{V}{Supported by MEXT and JSPS, Grants 18002006, 20540299, 18540281 and 26247032, the Daiko and Yamada Foundations (Japan)}
\item \Adef{W}{Supported by the Ministry of Science and Technology (Taiwan)}
\item \Adef{X}{Supported by the Israel Academy of Sciences and Humanities (Israel)}
\item \Adef{Y}{Supported by the Tomsk Polytechnic University within
the assignment of the Ministry of Science and Higher Education (Russia)}
\item \Adef{Z}{Supported by the National Science Foundation, Grant no. PHY-1506416 (USA)}
\item \Adef{a}{Supported by NCN, Grant 2017/26/M/ST2/00498 (Poland)}
\end{Authlist}

\clearpage
}
\setcounter{page}{1}

\section{Introduction}

The chiral-odd transversity quark distribution functions $h_1^q(x)$, hereafter referred to as transversity, were introduced as independent Parton Distribution Functions (PDFs) of the nucleon 
several decades ago~\cite{RaSo79,Baldracchini:1980uq,ArMe90,JaJi91}. Here, the superscript 
$q$ denotes the quark flavour and $x$ is the Bjorken variable. 
Several experimental approaches were proposed to access transversity 
in Semi-Inclusive measurements of Deep Inelastic Scattering (SIDIS) off transversely polarised nucleons. 

Two of these approaches, the measurements of 
Collins asymmetries~\cite{Airapetian:2004tw,Airapetian:2010ds,Alekseev:2010rw,Adolph:2012sn} and of
azimuthal asymmetries of hadron pairs produced on transversely polarised protons~\cite{Airapetian:2008sk,Adolph:2012nw,Adolph:2014fjw}, provided convincing evidence that transversity is indeed accessible experimentally.
For $u$- and $d$-quarks, transversity
was found to be different from zero at large $x$, where $h_1^u(x)$ and $h_1^d(x)$ are almost of the same size but opposite in sign, while $h_1^{\bar{u}}$ 
and $h_1^{\bar{d}}$ 
were found compatible with zero (see, e.g.,~Ref.~\cite{Martin:2014wua}). However, the uncertainties for the $d$($\bar{d}$)-quark are
about a factor of 3(2) larger than the uncertainties for the
$u$($\bar{u}$)-quark,
due to the unbalance of the existing proton and deuteron data.

A third approach, independent from the previous two, is the SIDIS measurement of the polarisation of baryons produced in the 
process  $\ell {\rm p}^\uparrow \to \ell {\rm B}^\uparrow {\rm X}$, where $\ell$ denotes a lepton, ${\rm p}^\uparrow$  a transversely polarised target proton and ${\rm B}$ a baryon~\cite{Baldracchini:1980uq,Artru:1990wq,Jaffe:1996wp,Anselmino:2003wu}. In the one-photon-exchange approximation, the hard interaction is $\gamma^* {\rm q}^\uparrow \to {\rm q}^{\prime \uparrow} $. When the virtual photon $\gamma^*$ interacts with a transversely polarised quark ${\rm q}$, the 
struck quark ${\rm q}^\prime$ has a certain probability to transfer
a fraction of the initial transverse polarisation to the final-state baryon. 
Thus a measurement of the polarisation of the final-state baryon along the 
spin direction of the outgoing quark allows 
access to transversity~\cite{Anselmino:2000ga,Barone:2003fy}. 

Among all baryons, $\Lambda$($\bar{\Lambda}$) hyperons are most suited to polarimetry studies due to their self-analysing weak decay into charged hadrons, $\Lambda \to {\rm p} \pi^-$ ($\bar{\Lambda} \to \bar{\rm p} \pi^+$), which occurs with a branching ratio $BR=63.9$\%. The polarisation $P_{\Lambda (\bar\Lambda)}$ is accessible through the modulation of the angular distribution
of the decay protons (antiprotons):

\begin{equation}
  \frac{\textrm{d}N_{\rm p(\bar{p})}}{\textrm{d} \cos \theta} \propto 1 + \alpha_{\Lambda (\bar\Lambda)}  P_{\Lambda (\bar\Lambda)}  \cos \theta \quad,
\label{meas_self_decay}
\end{equation}
where $\theta$ is the proton (antiproton) emission angle with respect to the polarisation axis of the fragmenting quark
in the $\Lambda (\bar\Lambda)$  rest frame and $\alpha_{\Lambda (\bar\Lambda)}$ is the weak decay constant. For the analysis presented in this paper, we 
use the most recent values of $\alpha_{\Lambda(\bar{\Lambda})}$~\cite{Ablikim:2018zay},
i.e., $\alpha_{\Lambda} = 0.750 \pm 0.009$ and $\alpha_{\bar{\Lambda}} = -0.758 \pm 0.010$.

\begin{figure}[t!]
\begin{center}
\includegraphics[width=0.4\textwidth]{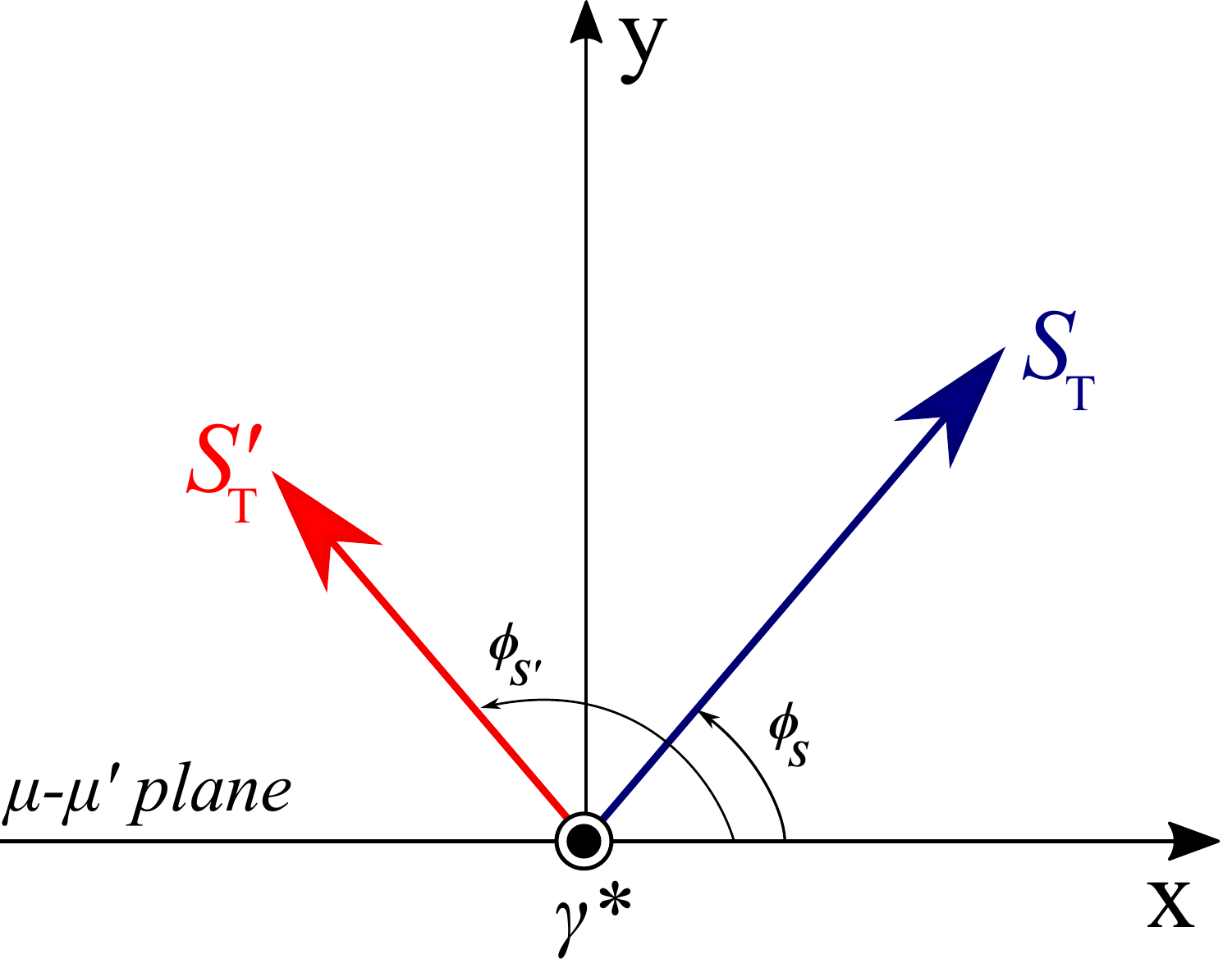}
\caption{Definition of the reference axes: The initial ($S_{\rm T}$) and final ($S_{\rm T}'$) transverse quark spin-polarisation vectors are shown with respect to the $\mu$ - $\mu'$ scattering plane.
}  
\label{coordinate2}
\end{center}
\end{figure}

As 
polarisation axis 
to access transversity 
we use the same that was used in QED calculations~\cite{Kotzinian:1994dv} 
for $\gamma^*$ absorption. Accordingly, the components of the quark spins in initial ($S_{\rm T}$) and final ($S'_{\rm T}$) state in the $\gamma^*$-nucleon system are connected by 
\begin{equation}
\label{gamma_abs}
S_{\rm T,x}' = - D_{\textrm{NN}} S_{\rm T,x} \, \qquad {\rm and} \qquad S_{\rm T,y}' =  D_{\textrm{NN}} S_{\rm T,y},
\end{equation}
where as z-axis the virtual-photon direction is taken
and as 
y-axis 
the normal to the lepton scattering (xz) plane (see Fig.~\ref{coordinate2}). The virtual-photon depolarisation factor $D_{\textrm{NN}}(y) = 2(1-y) / (1 + (1-y)^2)$ depends on $y$, the fraction of the initial lepton energy carried by the virtual photon in the target rest frame.
The polarisation direction $S'_{\rm T}$ of the fragmenting quark is obtained as the reflection of the initial quark polarisation $S_{\rm T}$ with respect to the y-axis. As this direction is uncorrelated with the normal to the $\Lambda$ production plane, no contribution of the spontaneous polarisation~\cite{PhysRevLett.36.1113} is expected 
after integration over the azimuthal angle of this plane. Note that this is independent of the polarisation reversal described in Sec.~\ref{sec:dr}.
For the same reason, also the twist-3 terms related to the beam polarisation, which could contribute to the observed $\Lambda$ polarisation along the $S'_{\rm T}$ direction (see e.g. Eqs. 100-102 in Ref.~\cite{Mulders:1995dh}), are expected to average to zero.

In the collinear approximation, where the intrinsic transverse momentum of the struck quark is assumed to be negligible, and in the current fragmentation region the expression for the  transversity-induced $\Lambda (\bar{\Lambda})$ polarisation integrated over the hadron transverse momentum $p_{\rm T}$ reads~\cite{Anselmino:2000ga}:

\begin{equation}\label{pt}
\begin{split}
P_{\Lambda (\bar{\Lambda})}(x,z,Q^2) & = \frac{{\rm d}\sigma^{\ell {\rm p}^{\uparrow} \rightarrow \ell^{\prime} \Lambda (\bar{\Lambda})^{\uparrow} {\rm X}} 
- {\rm d}\sigma^{\ell {\rm p}^{\uparrow} \rightarrow \ell^{\prime} \Lambda (\bar{\Lambda})^{\downarrow}{\rm X}}}
{{\rm d}\sigma^{\ell {\rm p}^{\uparrow} \rightarrow \ell^{\prime} \Lambda (\bar{\Lambda})^{\uparrow}{\rm X}}  
+ {\rm d}\sigma^{\ell {\rm p}^{\uparrow} \rightarrow \ell^{\prime}\Lambda (\bar{\Lambda})^{\downarrow}{\rm X}}} \\
& = f P_{\rm T}D_{\rm NN}(y) \frac{\sum_{q} e_{q}^{2}h_1^q(x,Q^2) H_{1,q}^{\Lambda (\bar{\Lambda})}(z,Q^2)}{\sum_{q} e_{q}^{2}f_1^q(x,Q^2) D_{1,q}^{\Lambda (\bar{\Lambda})}(z,Q^2)}.  
\end{split}
\end{equation}

Here, $Q^2$ is the photon virtuality and $z$ the fraction of the virtual photon energy carried by the $\Lambda (\bar{\Lambda})$ hyperon in the target rest frame; $P_{\rm T}$ is the target polarisation and $f$ the target dilution factor representing the fraction of nucleons effectively polarised in the target. The sums in Eq.(\ref{pt})  run over all quark and antiquark flavours. The transversity distribution functions $h_1^q(x,Q^2)$ appear coupled to the chiral-odd fragmentation functions 
$H_{1,q}^{\Lambda (\bar{\Lambda})}(z,Q^2)$ 
that describe the spin transfer from the struck quark to the $\Lambda (\bar{\Lambda})$ hyperon:
\begin{equation}
H_{1,q}^{\Lambda (\bar{\Lambda})}(z,Q^2)  = D_{1 ,   q^\uparrow}^{\Lambda (\bar{\Lambda})^\uparrow}(z,Q^2) - D_{1 , q^\uparrow}^{\Lambda (\bar{\Lambda})^\downarrow}(z,Q^2) \, .
\end{equation}  
The up and down arrows indicate the polarisation directions for the 
$\Lambda (\bar{\Lambda})$ along 
the $S'_{\rm T}$ axis. The polarisation-independent fragmentation functions
$D_{1,q}^{\Lambda (\bar{\Lambda})}(z,Q^2)$ are
 given by
\begin{equation}
D_{1,q}^{\Lambda (\bar{\Lambda})}(z,Q^2)  = D_{1 ,   q^\uparrow}^{\Lambda (\bar{\Lambda})^\uparrow}(z,Q^2) + D_{1 , q^\uparrow}^{\Lambda (\bar{\Lambda})^\downarrow}(z,Q^2) \, .
\end{equation}  
Evidently, this approach gives access to transversity only if at least a part of the quark spin is transferred to the final state hadron, i.e.~if $H_{1,q}^{\Lambda (\bar{\Lambda})}(z,Q^2)\neq 0$.
Alternatively, once transversity is known, $P_{\Lambda (\bar{\Lambda})}$ can be used to shed light on the size of the transverse-spin-dependent quark fragmentation function.

In general, $ P_{\Lambda (\bar\Lambda)}$ is not directly accessible from experimental data, as the detector acceptance distorts the angular distributions. Therefore, the measured angular distributions become

\begin{equation}
\frac{\textrm{d}N_{\rm p(\bar{p})}}{\textrm{d} \cos \theta } \propto \left (1 + \alpha_{\Lambda (\bar\Lambda)}  P_{\Lambda (\bar\Lambda)} \cos \theta \right ) \cdot A(\theta) \, ,
\label{def_self}
\end{equation}
where $A(\theta)$ is the detector acceptance depending on $\theta$, which generally would have to be studied via detailed Monte Carlo simulations. 
However, in the COMPASS experiment~\cite{Abbon:2007pq} 
the specific target setup offers the unique opportunity to measure the transversity-induced polarisation avoiding acceptance corrections (see Sec.~\ref{sec:dr}).

The analysis presented here was performed using the data collected by COMPASS in 2007 and 2010 with a 160 GeV/$c$ longitudinally polarised muon beam from the CERN SPS and a transversely polarised NH$_3$ target with proton polarisation $\langle P_{\rm T} \rangle = 0.80$ and dilution factor $\langle f \rangle = 0.15$. 
In an earlier analysis, the 
$\Lambda(\bar{\Lambda})$ polarisation from the 2002-2004 data with a transversely polarised deuteron target~\cite{Ferrero} was found to be compatible with zero, as expected from the cancellation of $u$ and $d$ quark transversity (see Sec.~\ref{deuteron_case}). This measurement, however, suffered from limitations in statistical power and in spectrometer acceptance and from the lack of particle identification for a  part of the data set. In this respect, the upcoming 2021/2022 run 
using a transversely polarised deuteron target~\cite{Friedrich:2286954} will be of great importance in drawing more definite conclusions.

\section{Data selection and available statistics}

In the data analysis, events  are selected if they have 
at least one primary vertex, defined as the intersection point 
of a beam track, the scattered muon track, and other possible outgoing 
tracks. 
The primary vertex is required to be inside a target cell. 
The  target consists of three cylindrical cells with 4~cm diameter,
a central one of 60~cm
and two outer ones of 30~cm length, 
each separated by 5~cm.
Consecutive cells are polarised in opposite directions, so that data 
with both spin directions are recorded at the same time~\cite{Alekseev:2010rw}.
The extrapolated beam track is required
to traverse all
three target cells to ensure equal muon flux through the full target.
Events originating from deep inelastic scattering are selected
by requiring $Q^2>1$ (GeV/$c$)$^2$. For the 
invariant mass of the final state 
produced in the interaction of virtual-photon and nucleon, 
$W> 5$ GeV/$c^2$ 
is required to avoid the region of exclusive resonance production. 
Furthermore the constraints  $0.003<x<0.7$ and $0.1<y<0.9$ are applied.
Here, the upper limit in $x$  
avoids a region of low statistics, 
and in $y$ the 
limits avoid
large radiative corrections
and
contamination from final-state pion
decay (upper limit)
and
warrant a good determination
of $y$ (lower limit).\\
The $\Lambda$ and $\bar{\Lambda}$ 
reconstruction is based on the detection of their decay products that originate from a
decay vertex ($V^0$) downstream of the production vertex, which is not connected to the latter by charged tracks. Due to the long $\Lambda$ lifetime, $\tau = (2.632 \pm 0.020) \cdot 10^{-10}$ s, both vertices can be well separated. Exactly two oppositely charged hadrons with  momentum larger than 1 GeV/$c$ are required to  originate from the decay vertex; the
reconstructed momentum vector for such a hadron pair is required to be aligned with the vector linking the production and the decay vertices within a collinearity angle $\theta_{\textrm{coll}} \leq $ 7 mrad.
In order to suppress background from photon conversion $\gamma\rightarrow e^+e^-$, the transverse momentum $p_{\perp}$ of each hadron, calculated with respect to the line-of-flight of the hadron pair in its rest frame,  has to be larger than 23 MeV/$c$.

Particle identification is performed using the 
RICH detector. In order to limit the ambiguity between $\Lambda$($\bar{\Lambda}$) hyperons and
$K_{\rm s}^0$ mesons decaying into $\pi^+\pi^-$, it is necessary to ensure that the positive (negative) daughter particle is a proton (antiproton). However, a direct identification would  drastically reduce the available statistics due to the high Cherenkov threshold for protons of about 20 GeV/$c$ for  the radiator gas used (C$_{4}$F$_{10}$). 
Therefore, assuming one charged track as negative (positive) pion, the corresponding positive (negative) track is considered to be a proton (anti-proton) unless it is identified as
positive (negative) electron, pion or kaon.
The particle identification procedure is the same as it was used in previous analyses \cite{Abbon:2011zza}. It is based on the calculation of the maximum likelihood $\mathcal{L}$ for four mass hypotheses ($e$, $K$, $\pi$, $p$) and for the background, given the number of collected Cherenkov photons. In order to attribute a mass hypothesis $M$ to a particle, $\mathcal{L}_M$ is requested to be the highest and its ratio to the background hypothesis to be larger than an optimised threshold. This approach is applied to particles with momentum up to 50 GeV/$c$, a value at which
pion/kaon separation becomes difficult. Beyond this limit, the highest likelihood is required not to be the one associated to the pion or kaon mass hypothesis.

\begin{center}
\begin{figure}[t!]
    \centering
    \includegraphics[width=0.65\textwidth]{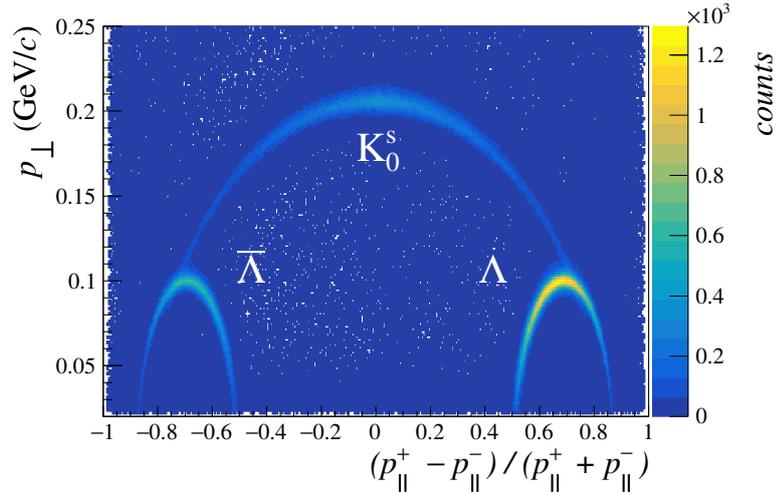}
    \caption{Armenteros-Podolanski plot. 
    } \label{fig:armenteros}
\end{figure}
\end{center}

\begin{center}
\begin{figure}[b!]
   \centering
     \includegraphics[width=0.48\textwidth]{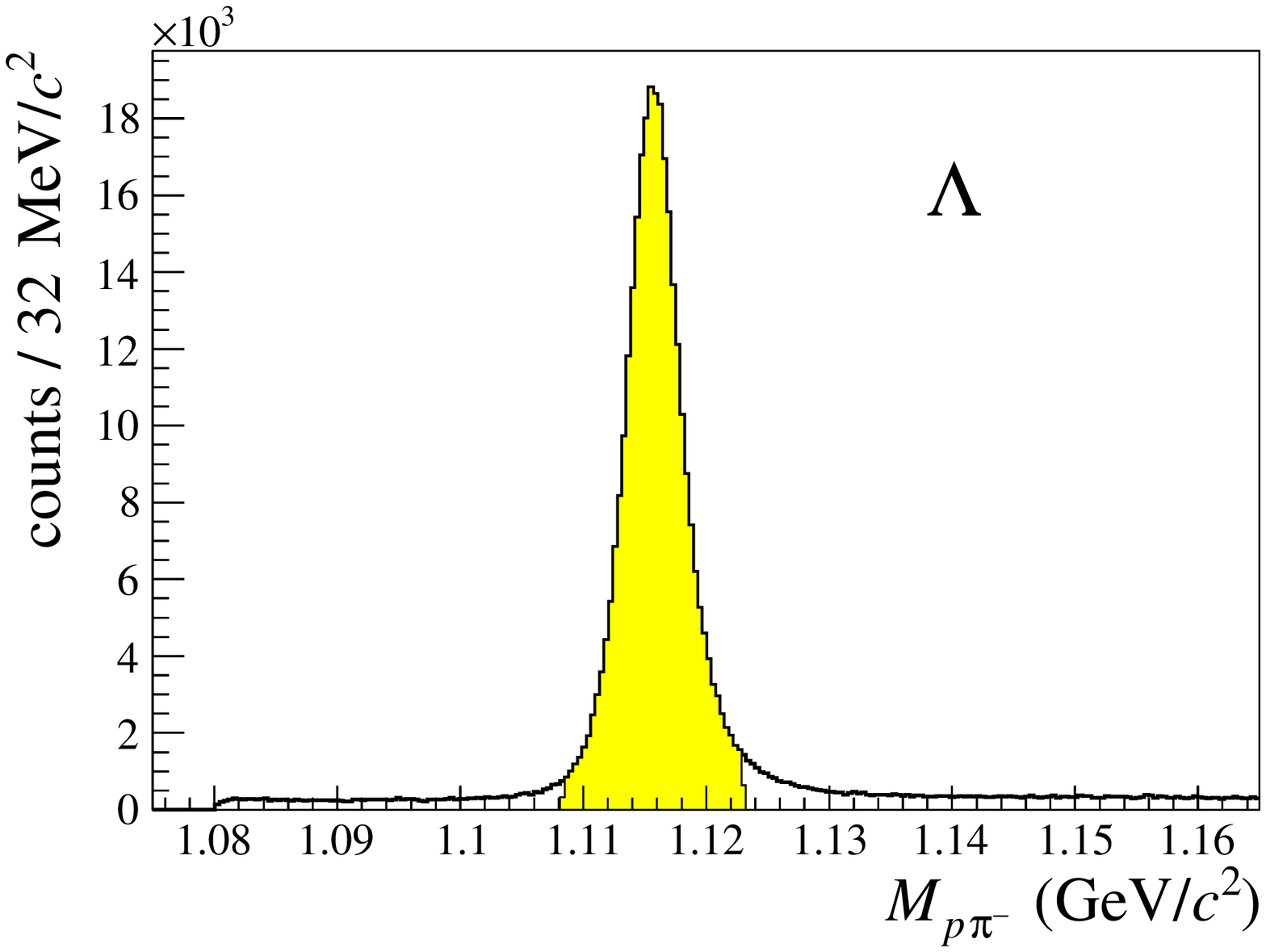}
    \includegraphics[width=0.48\textwidth]{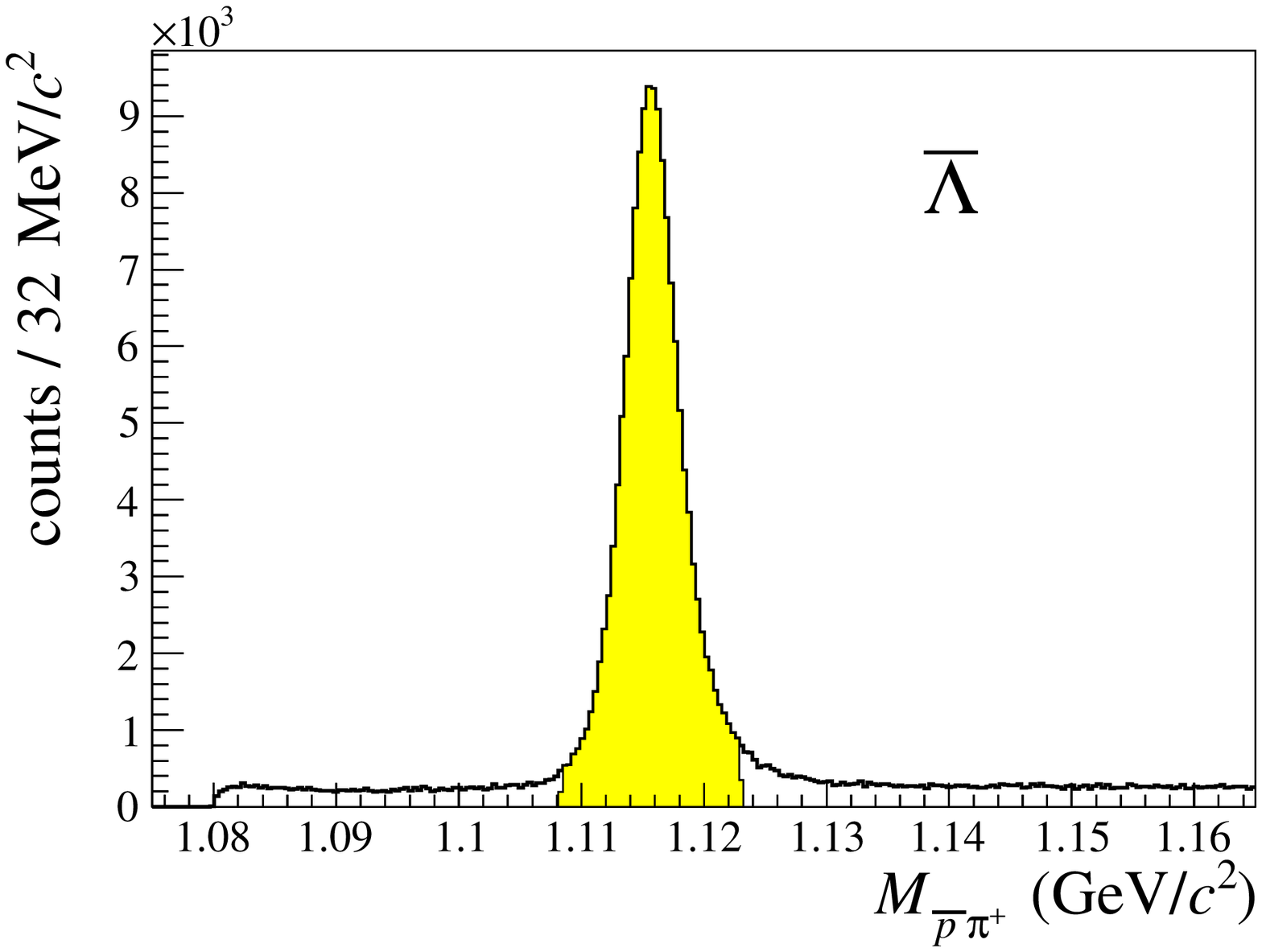}
   \caption{Invariant mass spectra of $\Lambda$ (left) and $\bar{\Lambda}$ (right) after all selection steps.} \label{fig:armenteroslr}
\end{figure}
\end{center}

The Armenteros-Podolanski plot \cite{Armenteros1, Armenteros2} obtained after all aforementioned selection steps is shown in Fig.~\ref{fig:armenteros}. The remaining $K_{\rm s}^0$ contribution to the selected sample is visible as the symmetric arc, while a selection of the left and right halves of the 
figure allows to separate $\bar{\Lambda}$
(on the left) from $\Lambda$ hyperons (on the right), based on the sign of the longitudinal momentum asymmetry $(p_{\parallel}^+-p_{\parallel}^-) / (p_{\parallel}^++p_{\parallel}^-)$.
Here, $p_{\parallel}^+$($p_{\parallel}^-$) indicates the longitudinal momentum of the positive  (negative) decay particle in the hyperon rest frame
with respect to the  $\Lambda$($\bar{\Lambda})$ line of flight. 
In Fig.~\ref{fig:armenteroslr} the $\Lambda$ and $\bar{\Lambda}$ invariant mass spectra corresponding to these selections are shown. Here, only the $K_{\rm s}^0$ in the crossing regions of the  $K_{\rm s}^0$ and $\Lambda$($\bar{\Lambda}$) arcs contribute to the background.  These invariant mass spectra 
are fitted with a superposition of a Gaussian function and a constant term using the PDG value for the $\Lambda$ mass \cite{Tanabashi:2018oca}.
The background is evaluated with the sideband method considering two equally wide intervals on the left and on the right of the mass peak. 
Finally, 
hyperons are selected within a~$\pm 3\sigma$ range from the peak, where $\sigma = 2.45 $ MeV/$c^2$ 
is obtained using all data
shown in Fig.~\ref{fig:armenteroslr}. Depending on the chosen kinematic bin, the signal-over-background ratio ranges from 5.7 to 54.9. The total statistics after background subtraction  
are given in Tab.~\ref{table:stat}.

A significant fraction of $\Lambda$ and $\bar{\Lambda}$ particles originates from the decay of heavier hyperons. Using the event generator
LEPTO  
based on the Lund string model \cite{Ingelman:1996mq}, 63\% of the  $\Lambda$
and 68\% of the $\bar{\Lambda}$ hyperons produced in the 
COMPASS kinematic regime are estimated to originate
from direct string fragmentation~\cite{Adolph:2013dhv}. 
In this analysis, the $\Lambda$ and $\bar{\Lambda}$ hyperons coming from indirect production cannot be separated from those coming from direct production. Their contribution is not taken into account as a  systematic uncertainty in the analysis, although it could dilute a  possible polarisation signal.

\begin{center}
\begin{table}[t!]
   \caption{Available statistics for $\Lambda$ and $\bar{\Lambda}$ hyperons, after background subtraction, for years 2007 and 2010 and for their sum.}
    \centering
    \begin{tabular}{ c c c  }
 year & $\Lambda$ & $\bar{\Lambda}$\\ \hline 
 2007 & ~~95\,125 $\pm$ 315 & ~~44\,911 $\pm$ 227\\  
 2010 & 201\,421 $\pm$ 466 & ~~99\,552 $\pm$ 336 \\ \hline
 total & 296\,546 $\pm$ 562 & 144\,463 $\pm$ 405\\ \hline
\end{tabular}
\label{table:stat}
\end{table}
\end{center}

\section{Extraction method and results for $\Lambda(\bar{\Lambda})$ polarisation}
\label{sec:dr}

For this analysis, as for all target spin asymmetries measured at COMPASS, systematic effects are minimised due to the unique target configuration described 
at the beginning of the previous section
and to the fact that the data taking is divided into periods, each consisting of two subperiods in which data are taken with reversed polarisation orientation in each target cell.

As the transversity-induced $\Lambda (\bar{\Lambda})$ polarisation is to be measured
along the spin direction of the fragmenting quark, this reference axis has to be determined
on an event-by-event basis. The initial-quark spin is assumed to be aligned with the nucleon spin and is thus vertical in the laboratory frame. Its transverse component is rotated by an azimuthal angle $\phi_{S}$ in the $\gamma^*$-nucleon system (Fig.~\ref{coordinate2}). 
As described above, the spin direction of the quark after the interaction with the virtual photon is obtained by reflecting it with respect to the normal to the lepton scattering plane~\cite{Barone:2003fy,Kotzinian:1994dv,Baum:1996yv},  $\phi_{ S^\prime}=\pi -\phi_{S}$. In the present 
analysis we 
determine
the
$\Lambda$($\bar{\Lambda}$) polarisation along this direction.

The number of $\Lambda$($\bar{\Lambda}$) hyperons emitting a proton (antiproton) in a given $\cos\theta$ range from a given target cell with a given direction of the target polarisation can be expressed as

\begin{equation}\label{N}
    \mathcal{N}_{\Lambda (\bar{\Lambda}), i}^{\{\prime\}}(\cos\theta) = \Phi_i^{\{\prime\}} \ \rho_i^{\{\prime\}} \ \bar{\sigma}_{\Lambda (\bar{\Lambda})} \ (1 \varpm \alpha_{\Lambda (\bar\Lambda)} P_{\Lambda (\bar{\Lambda})} \cos(\theta+(i-1)\pi)) \ A_i^{\{\prime\}}(\cos\theta) \, .
\end{equation}
Here, $i=1,2$ indicates the central or outer cells, respectively, $\Phi_i^{\{\prime\}}$ denotes the muon flux, $\rho_i^{\{\prime\}}$ the number of nucleons per unit area,
and $\bar{\sigma}_{\Lambda(\bar{\Lambda})}$ is the cross section for the production of $\Lambda$($\bar{\Lambda}$) hyperons. The
acceptance term $A_i^{\{\prime\}}(\cos\theta)$ includes both geometrical acceptance, which is
slightly different for each of the three target cells, and \mbox{spectrometer} efficiency. Primed quantities refer to
data taken in 
subperiods after target polarisation reversal. After background subtraction, the four equations of Eq.(\ref{N}) are combined to form a 
double ratio 
\begin{equation}\label{DR_def}
     \varepsilon_{\Lambda (\bar{\Lambda})}(\cos\theta) = \frac{\mathcal{N}_{\Lambda (\bar{\Lambda}),1}^{~}(\cos\theta) \mathcal{N}_{\Lambda (\bar{\Lambda}),2}^\prime (\cos\theta)}{\mathcal{N}_{\Lambda (\bar{\Lambda}),1}^\prime (\cos\theta) \mathcal{N}_{\Lambda (\bar{\Lambda}),2}^{~}(\cos\theta)} ~~.
\end{equation}
As described in Refs.~\cite{Alexakhin_2005,Ageev:2006da}, the acceptances cancel in this expression as long as 
in each $\cos\theta$ bin  
the acceptance ratios for the target cells after polarisation reversal are equal to those before, which is
a reasonable assumption for the given setup.
As described above, equal muon flux in all three target cells is maintained by the event selection, so that also the flux cancels in Eq.(\ref{DR_def}). For small values of the $\Lambda (\bar{\Lambda})$ polarisation it then becomes:
\begin{equation}
     \varepsilon_{\Lambda (\bar{\Lambda})}(\cos\theta)  \approx 1 + 4\alpha_{\Lambda (\bar{\Lambda})} P_{\Lambda (\bar{\Lambda})} \cos\theta.
\end{equation}

In each kinematic bin in $x$, $z$ or $p_{\rm T}$, the data sample is divided into eight $\cos\theta$ bins. This set of eight $\varepsilon_j$ values is then fitted with the linear function $f=p_0(1+p_1\cos\theta)$,  
so that $P_{\Lambda (\bar{\Lambda})}$ 
is obtained as $P_{\Lambda (\bar{\Lambda})} = p_1 / (4\alpha_{\Lambda (\bar{\Lambda})})$.\\

The transversity-induced polarisation is measured in the full phase-space and in the following regions:

\begin{itemize}
\setlength\itemsep{0em}
    \item current fragmentation: $z \geq 0.2$ and Feynman variable $x_{\rm F} > 0$;
    \item target fragmentation: $z<0.2$ or $x_{\rm F}<0$;
    \item high $x$: $x \geq 0.032$;
    \item low $x$: $x<0.032$;
    \item high $p_{\rm T}$: $p_{\rm T} \geq 0.5$ GeV/$c$;
    \item low $p_{\rm T}$: $p_{\rm T}<0.5$ GeV/$c$.
\end{itemize}

In each of these regions, the data is
scrutinised for possible systematic biases. The two main sources of systematic uncertainties are period compatibility and false $\Lambda (\bar{\Lambda})$ polarisations.
The former are
evaluated by comparing the results from the various periods of data taking, while 
the latter are evaluated by 
reshuffling the double ratio from Eq.(\ref{DR_def}) as 
$(\mathcal{N}_{\Lambda (\bar{\Lambda}),1}^{~} \mathcal{N}_{\Lambda (\bar{\Lambda}),2}^{~})/(\mathcal{N}_{\Lambda (\bar{\Lambda}),1}^\prime \mathcal{N}_{\Lambda (\bar{\Lambda}),2}^\prime)
$, so that transversity-induced $\Lambda (\bar{\Lambda})$ polarisations cancel.
Effects of residual acceptance variations are proven to be negligible by evaluating the $K_{\rm s}^0$ polarisation 
that is found to be compatible with zero as expected. 
In addition, $P_{\Lambda(\bar{\Lambda})}$ is measured assuming
the central cell split into two halves, thus creating two data samples by combining each half with one of the outer cells. Again, effects of acceptance variation are found to be negligible.
A scale uncertainty of about 7.5\% contributes to the overall systematics due to the uncertainty on the weak decay constant $\alpha$ (2\%) and on the dilution and polarisation factors $f$ and $P_{\rm T}$ (5\% overall). In general, $\sigma_{\textrm{syst}} < 0.85~\sigma_{\textrm{stat}}$.
If the intrinsic transverse momentum $k_{\rm T}$ of the initial quark is not assumed to be zero, as it is done in the collinear approximation, the expression for $P_\Lambda$ involves 
several other terms \cite{Boer:1999uu} dependent on
the azimuthal angle  $\phi$ of the $\Lambda(\bar{\Lambda})$ hyperon and on 
$\phi_S$, which should be taken into consideration
in case of a non-flat acceptance in terms of these angles. It was checked that the acceptance is flat over the azimuthal angle  $\phi$, so that after integration the contribution of $\phi$-dependent terms vanishes. A remaining  contribution  of $\cos2\phi_S$-dependent terms can not a priori be excluded because the acceptance in $\phi_S$ is not flat.
It was checked that the measured polarisations extracted for positive and negative values of $\cos2\phi_S$ are consistent within statistical uncertainties, so that the contribution of $\cos2\phi_S$-dependent terms can be neglected.

\begin{center}
\begin{figure}[!ht]
    \centering
    \includegraphics[width=0.85\textwidth]{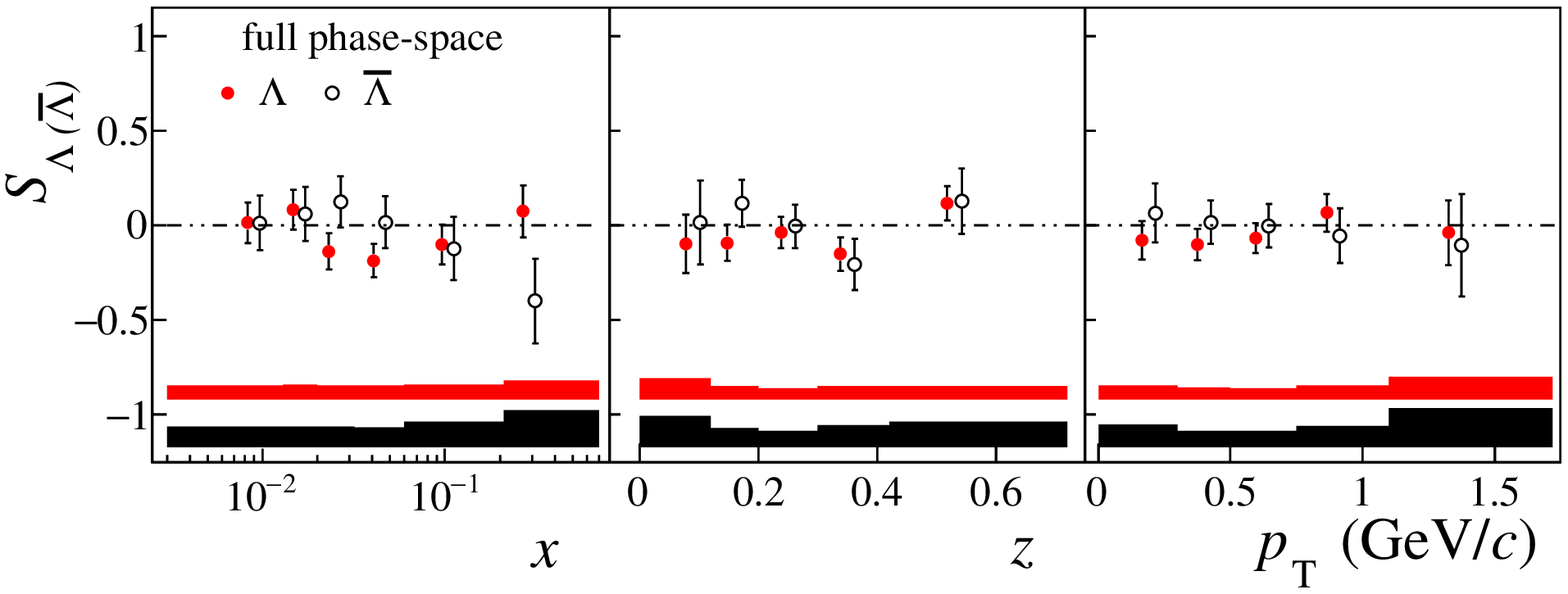}
    \includegraphics[width=0.85\textwidth]{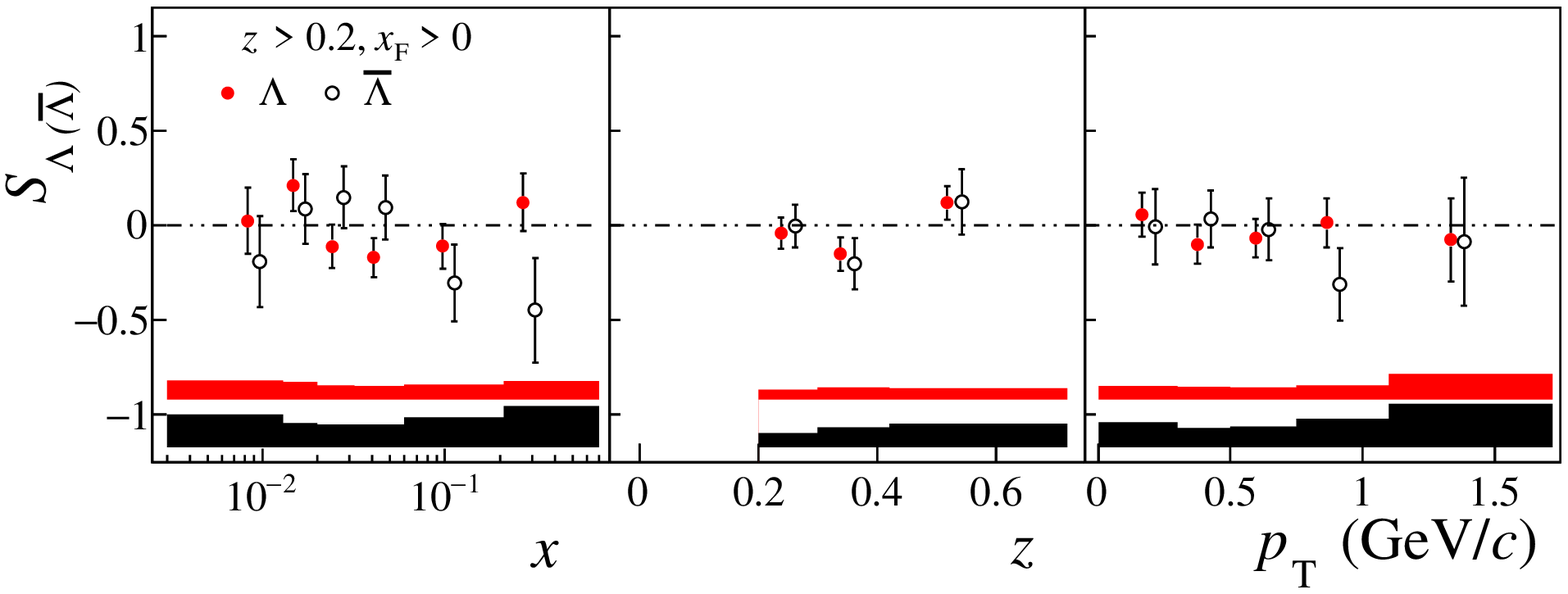}
    \caption{Spin transfer $S_{\Lambda(\bar{\Lambda})}$ for the full phase-space (top) and for the current fragmentation region (bottom), as a function of $x$, $z$ and $p_{\rm T}$ . The bands show the systematic uncertainties, while the error bars represent statistical uncertainties. The values in $x$, $z$ and $p_{\rm T}$ are staggered for clarity.}
        \label{fig:all_pol}
\end{figure}
\end{center}

In Fig.~\ref{fig:all_pol}, the results from the full phase-space and for the current fragmentation region
are presented in terms of the spin transfer
\begin{equation}\label{st}
    S_{\Lambda (\bar{\Lambda})} = \frac{P_{\Lambda (\bar{\Lambda})}}{fP_{\rm T} D_{\rm NN}(y)},
\end{equation}
by definition ranging from -1 to 1. The corresponding numerical values are given in the Appendix. The full set of data for all
selections can be found on HEPDATA \cite{Maguire:2017ypu}.

\section{Interpretation of the results and 
predictions for future measurements}

The polarisations shown in Fig.~\ref{fig:all_pol} are compatible with zero within the experimental uncertainties in all 
studied kinematic regions.
From this result, applying different 
hypotheses, some conclusions 
will be drawn below
on the size of the 
fragmentation function $H_{1,u}^{\Lambda}(z,Q^2)$ as well as on the 
strange quark transversity distribution  $h_1^s(x,Q^2)$.\\ 

Following Eq.(\ref{pt}) and Eq.(\ref{st}), in the current fragmentation region the spin transfer $S_{\Lambda(\bar{\Lambda})}$  reads

\begin{equation}\label{stranfer}
    S_{\Lambda (\bar{\Lambda})} = \frac{\sum_{q}e_q^2 h_1^{q} H_{1,q}^{{\Lambda (\bar{\Lambda})}}}{\sum_{q}e_q^2 f_1^{q} D_{1,q}^{{\Lambda (\bar{\Lambda})}}},
\end{equation}
where the dependences on $x$, $z$ and $Q^2$ are omitted for simplicity. \\

\subsection{Interpretation of the measured $\bar{\Lambda}$ polarisation}

Considering the case of $\bar{\Lambda}$ hyperons, 
the favoured fragmentation functions 
$H_{1,\bar{u}}^{\bar{\Lambda}}$,
$H_{1,\bar{d}}^{\bar{\Lambda}}$ and $H_{1,\bar{s}}^{\bar{\Lambda}}$ only appear in combination 
with the sea-quarks $\bar{u}$, $\bar{d}$ and $\bar{s}$.
As $h_1^{\bar{s}} \approx 0$ can be assumed in analogy to $h_1^{\bar{u}}$ and  $h_1^{\bar{d}}$, transversity is coupled only to unfavoured fragmentation functions. Here  $H_{1,u}^{\bar{\Lambda}}$ and $H_{1,d}^{\bar{\Lambda}}$ dominate, as the $s$-quark contribution $h_1^{s} H_{1,s}^{\bar{\Lambda}}$ can be neglected because also $h_1^{s} $ is expected to be small. This yields

\begin{equation}
    \sum_{q}e_q^2 h_1^{q}H_{1,q}^{\bar{\Lambda}} \propto 4h_1^uH_{1,u}^{\bar{\Lambda}} + h_1^dH_{1,d}^{\bar{\Lambda}}
~~.
\end{equation}

The compatibility with zero  of the measured polarisation for $\bar{\Lambda}$ hyperons is in agreement with expectations 
based on calculations for the ratios of favoured to unfavoured fragmentation functions (see, e.g., Ref.~\cite{Ma:2003gd}). In these calculations, the unfavoured fragmentation functions are suppressed by a factor of about  $4$ to $5$
in the current fragmentation region at $z$ about $0.2$ and rapidly decrease further for increasing $z$.

\subsection{Interpretation of the measured $\Lambda$ polarisation}

Considering the case of
$\Lambda$ hyperons, 
one of the options suggested in e.g. Ref.~\cite{Ma:2003gd} is to
retain only the favoured combinations
($H_{1,u}^{\Lambda}$, $H_{1,d}^{\Lambda}$, $H_{1,s}^{\Lambda}$, $D_{1,u}^{\Lambda}$, $D_{1,d}^{\Lambda}$, $D_{1,s}^{\Lambda}$)
in both  numerator and  denominator,
resulting in:
\begin{equation}\label{stransfer1}
    S_{\Lambda} = \frac{4h_1^{u} H_{1,u}^{\Lambda} + h_1^{d} H_{1,d}^{\Lambda} + h_1^{s} H_{1,s}^{\Lambda}}{4f_1^{u} D_{1,u}^{\Lambda} + f_1^{d} D_{1,d}^{\Lambda} + f_1^{s} D_{1,s}^{\Lambda}}.
\end{equation}
Isospin symmetry requires $D_{1,d}^{\Lambda} = D_{1,u}^{\Lambda}$ and $H_{1,d}^{\Lambda}= H_{1,u}^{\Lambda}$. 
For the $s$-quark fragmentation functions,
it is often assumed that
$D_{1,s}^{\Lambda}$ is proportional to $D_{1,u}^{\Lambda}$ with  the proportionality constant $r$, which is the inverse of the 
strangeness suppression 
factor $\lambda =1/r$
\cite{PhysRevD.59.114012,PhysRevD.58.094014}. In Ref.~\cite{Yang:2002gh} its value is obtained from a fit of experimental baryon production data in $e^+e^-$ annihilation to be $\lambda_\Lambda=1/r=0.44$. With these simplifications, Eq.(\ref{stransfer1}) turns into

\begin{equation}
    S_{\Lambda} = \frac{\left[4h_1^u+h_1^d\right] H_{1,u}^{\Lambda} + h_1^sH_{1,s}^{\Lambda}}{\left[4f_1^u+f_1^d+r f_1^s\right]D_{1,u}^{\Lambda}}.
\end{equation}

The interpretation is now performed in three different scenarios.
When needed, we use the CTEQ5D PDFs \cite{Buckley:2014ana} 
for $f_1^q$,  calculated at the $x$ and $Q^2$ values of the data points, while the values of the  transversity functions for $u$ and $d$ quarks are obtained from the fit 
presented in Ref.~\cite{Martin:2014wua}.

\subsection*{i) Transversity is non-zero only for  
valence quarks in the nucleon }

If transversity is assumed non-vanishing only for valence quarks, $h_1^s$ can be neglected and the expression for the spin transfer to the $\Lambda$ further simplifies to:

\begin{equation}
    S_{\Lambda} = \frac{[4h_1^u+h_1^d]H_{1,u}^{\Lambda}}{[4f_1^u+f_1^d+rf_1^s]D_{1,u}^{\Lambda}}.
\end{equation}

When $S_{\Lambda}$ is now inspected only as a function of $x$, its dependence upon $z$, carried by the fragmentation functions, is integrated over. In a generic $x$ bin centered at $x^*$ it becomes

\begin{equation}
    S_{\Lambda}|_{x=x^*} = \frac{[4h_1^u(x^*)+h_1^d(x^*)] \int_{0.2}^{1.0} {\rm d}z H_{1,u}^{\Lambda}(z)}{[4f_1^u(x^*)+f_1^d(x^*)+rf_1^s(x^*)]\int_{0.2}^{1.0} {\rm d}z D_{1,u}^{\Lambda}(z)}.
\end{equation}

Thus the measurement of $S_{\Lambda}$ as a function of $x$ can be used to extract, in each bin of $x$, the ratio $\mathcal{R}$ of the $z$-integrated fragmentation functions $H_{1,u}^{\Lambda}$ and $D_{1,u}^{\Lambda}$:

\begin{equation}
\mathcal{R}(x^*)= \left. \frac{\int_{0.2}^{1.0} {\rm d}z H_{1,u}^{\Lambda}(z)}{\int_{0.2}^{1.0} {\rm d}z D_{1,u}^{\Lambda}(z)}  \right|_{x=x^*} = \frac{4f_1^u(x^*)+f_1^d(x^*)+r f_1^s(x^*)}{4h_1^u(x^*)+h_1^d(x^*) } S_{\Lambda}|_{x=x^*}.
\end{equation}
In Tab.~\ref{table:R} the mean values of $\mathcal{R}$ over the measured $x$ range are given for three 
different choices of
$r$. These values, which show a weak dependence on $r$, are all negative and still compatible with zero within the given uncertainties.

\begin{center}
\begin{table}[!h]
    \caption{Mean value $\langle \mathcal{R} \rangle$ of the ratio of the $z$-integrated fragmentation functions for three choices of $r$. The uncertainties are statistical only.}
    \centering
    \begin{tabular}{ c c  }\hline
 $r$ & $\langle \mathcal{R} \rangle \pm \sigma_{\langle\mathcal{R}\rangle}$ \\ \hline 
 2 &  -- 0.27 $\pm$ 0.55 \\  
 3 &  -- 0.27 $\pm$ 0.56 \\   
 4 &  -- 0.26 $\pm$ 0.57 \\ \hline
\end{tabular}
\label{table:R}
\end{table}

\end{center}

\subsection*{ii) $\Lambda$ polarisation is carried by the $s$ quark only}

\begin{center}
\begin{figure}[!h]
    \centering
    \includegraphics[width=0.7\textwidth]{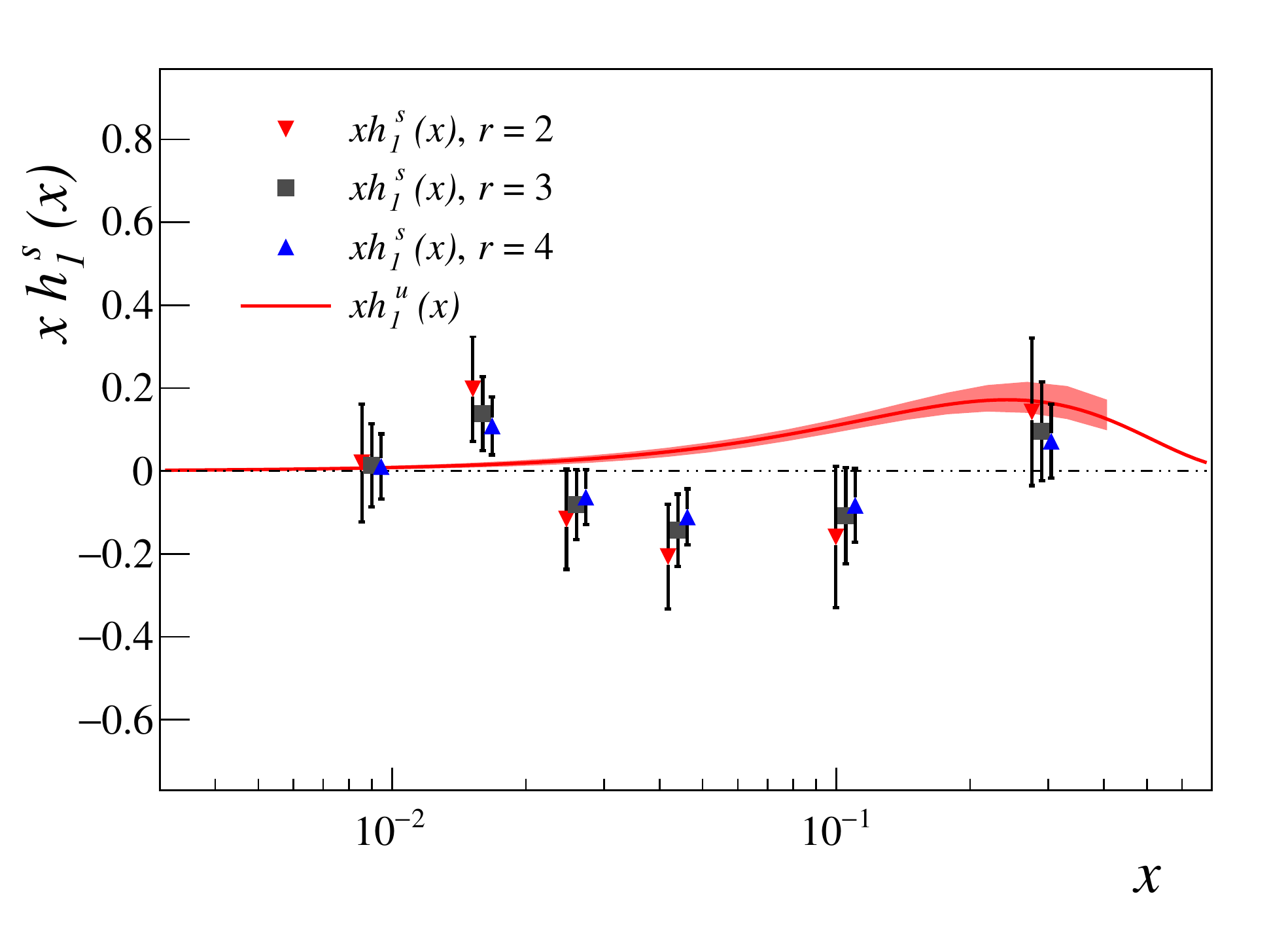}
    \caption{Extracted values of $xh_1^s(x)$ for the three options $r=2,3,4$. The $u$ quark transversity curve from Ref.~\cite{Martin:2014wua} is given for comparison. Only statistical uncertainties are shown and the $x$ values are staggered for clarity.}
        \label{fig:xh1s}

\end{figure}
    
\end{center}

Assuming instead that the polarisation is entirely carried by the $s$ quark, as in the SU(3) non-relativistic quark model, $H_{1,u}^{\Lambda}$ can be neglected. Moreover, as
suggested by Ref.\cite{Yang:2001sy}, 
$H_{1,s}^{\Lambda}$ can be approximated with $D_{1,s}^{\Lambda}$ for $z>0.2$, yielding

\begin{equation}
        S_{\Lambda} = \frac{h_1^s \, H_{1,s}^{\Lambda}}{\left[4f_1^u+f_1^d+ r f_1^s\right]\frac{1}{r}D_{1,s}^{\Lambda}} \approx \frac{r \, h_1^s}{4f_1^u+f_1^d+r f_1^s}\, ,
\end{equation}
so that $h_1^s$ can be extracted. In Fig.~\ref{fig:xh1s} the quantity $xh_1^s(x)$ is given for various choices of $r$ and compared to the fitted value and accuracy of the $xh_1^u(x)$ distribution \cite{Martin:2014wua}.
Again, only a weak dependence on $r$ is observed. Although the data suggest a negative sign of $h_1^s(x)$, they are not
precise enough to determine accurately $h_1^s(x)$ compared to the statistical precision of the $h_1^u(x)$ data.

\subsection*{iii) Polarised-$\Lambda$ production is described by a quark-diquark fragmentation model}

\begin{center}
\begin{figure}[t!]
    \centering
    \includegraphics[width=0.7\textwidth]{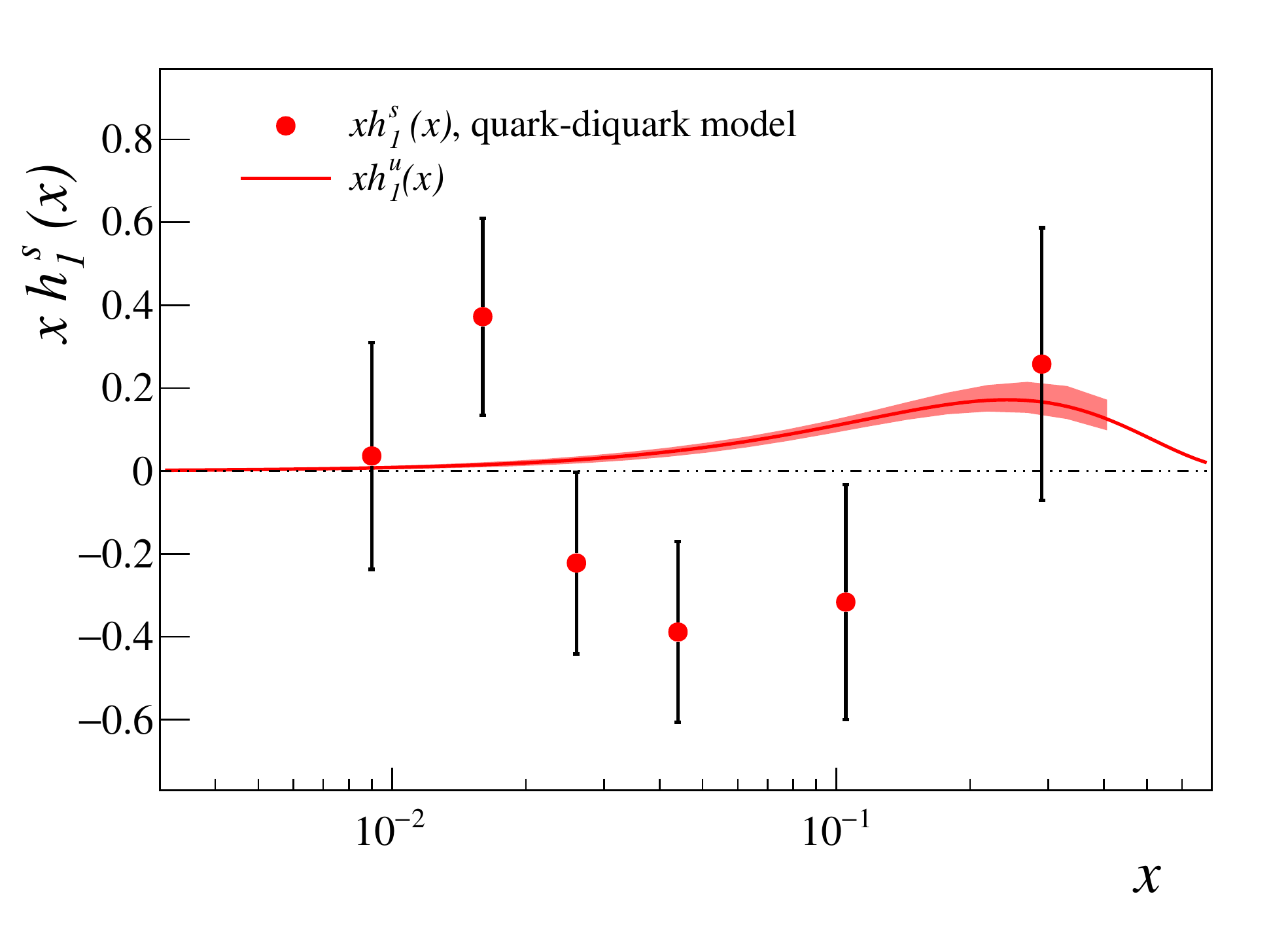}
    \caption{Extracted values of $xh_1^s(x)$ according to 
    a quark-diquark model \cite{Yang:2001sy, Jakob:1997wg}. The $u$ quark transversity curve from  Ref.~\cite{Martin:2014wua} is given for comparison. Only statistical uncertainties are shown.}
        \label{fig:xh1s_y}
\end{figure}
    
\end{center}

In the context of the quark-diquark model \cite{Yang:2001sy, Jakob:1997wg}, the fragmentation of an unpolarised valence quark $\rm{q}$ into a final-state hadron 
is accompanied by the emission of a diquark $D$, which can be in a scalar ($S$) or  vector ($V$) spin configuration. The probabilities $a_{\rm D}^{(q)}(z)$ associated to these two configurations are calculated in the model and enter the definition of the quark fragmentation function, which depends on $z$ and on the masses of the fragmenting quark,  the diquark and the produced hadron. Analogously, the fragmentation of a polarised quark is described through the probabilities $\hat{a}_{\rm D}^{(q)}(z)$. In the case of $\Lambda$ production, the unpolarised fragmentation function of the $s$ quark, $D_{1,s}^{\Lambda}$, is taken as reference and used to express 
all the other fragmentation functions  by introducing the flavour structure ratios $F_{\rm S}^{(u/s)}(z) = a_{\rm S}^{(u)}(z) / a_{\rm S}^{(s)}(z)$, $F_{\rm M}^{(u/s)}(z) = a_{\rm V}^{(u)}(z) / a_{\rm S}^{(s)}(z)$ and the spin-structure ratios $\hat{W}_{\rm D}^{q}(z) = \hat{a}_{\rm D}^{(q)}(z) / a_{\rm D}^{(q)}(z)$. The transversity-induced polarisation can thus be written as:
\begin{center}
  \begin{equation}\label{eq_diq}
        S_{\Lambda} = \frac{\left(4h_1^u+h_1^d\right) \cdot \frac{1}{4}\left[ \hat{W}_{\rm S}^{(u)}F_{\rm S}^{(u/s)} - \hat{W}_{\rm V}^{(u)}F_{\rm M}^{(u/s)}\right] +  h_1^s\hat{W}_{\rm S}^{(s)}}   {\left(4f_1^u+f_1^d\right) \cdot \frac{1}{4}\left[ F_{\rm S}^{(u/s)} +3 F_{\rm M}^{(u/s)}\right] +  f_1^s},
\end{equation}
\end{center}

where the $x$ and $z$ dependences are  omitted for clarity.
Information on $h_1^s$ can be obtained by integrating  
Eq.(\ref{eq_diq}) over $z$ in each $x$ bin. The values of $xh_1^s(x)$, as predicted by the 
quark-diquark model and based on the measured polarisation, are shown in Fig.~\ref{fig:xh1s_y}. The dependence of the final results on the mass of the diquark (containing or not the $s$ quark) was found negligible. \\
Again, as in scenario ii), the data suggest a negative sign of $h_1^s(x)$, but statistical uncertainties are even larger in this case. Improved data will be needed to determine $h_1^s(x)$ more accurately.

\subsection{Projections for future data taking with transversely polarised deuterons}
\label{deuteron_case}

The upcoming COMPASS run  aims at collecting new  precision SIDIS data using a polarised deuteron (LiD) target. The expected statistical uncertainties of the measured asymmetries are in the order of 60\% of those estimated for the proton data. Compared to the existing deuteron data taken with the early COMPASS setup, we expect an accuracy improvement
between a factor of two at small $x$ and a factor of five at large $x$~\cite{Friedrich:2286954}.
Some prospects for this measurement 
are described in the following. \\
The expression for the spin transfer, for $\Lambda$ production on a transversely-polarised deuteron target, 
reads:
\begin{equation} \label{eq_d}
    S_{\Lambda}^D = \frac{5(h_1^u+h_1^d)H_{1,u}^{\Lambda} + 2h_1^sH_{1,s}^{\Lambda}}{5(f_1^u+f_1^d)D_{1,u}^{\Lambda} + 2f_1^sD_{1,s}^{\Lambda}}.
\end{equation}

It is already known from earlier  COMPASS data that $h_1^d \approx - h_1^u$~\cite{Barone:2019yvn,Anselmino:2020vlp}.
The upcoming COMPASS run on a deuteron target will, in addition, allow us to measure with  high precision the quantity $h_1^u + h_1^d$.
Since the fragmentation function $H_{1,u}^{\Lambda}$ is expected to be smaller than the fragmentation  function $H_{1,s}^{\Lambda}$,
the numerator of
Eq.(\ref{eq_d}) will be dominated by the 
product $h_1^s  H_{1,s}^{\Lambda}$ if $h_1^s$ is of significant size. Therefore,
a new high statistics measurement of the transversity-induced $\Lambda$ polarisation
on a deuteron target from the upcoming data is  expected to be very sensitive to the  product $h_1^s  H_{1,s}^{\Lambda}$.

\section{Summary and outlook}

Using a transversely polarised proton target and a 160 GeV/$c$ muon beam, the transversity-induced polarisation along the 
spin axis of the
struck quark was measured by COMPASS for $\Lambda$ and $\bar{\Lambda}$ hyperons. 
While considered to be an excellent channel to access transversity, the results were found to be compatible with zero in all 
studied kinematic regions.\\ 
 The statistical uncertainty on the measured polarisation is still large,  despite the fact that all COMPASS data on a transversely  polarised proton target  were used, which are  the only existing world data suitable for this measurement.
Nevertheless, some information could be deduced from the existing data.\\
Under the hypothesis that transversity 
is non-vanishing only for valence quarks, 
the data were used to investigate the ratio of  $z$-integrated polarised to
unpolarised fragmentation functions. The results  
indicate a negative ratio, although compatible with zero due to the large uncertainties.
If instead a non-relativistic SU(3) quark model or a quark-diquark model is
considered, some information can be derived on the transversity distribution for the $s$ quark. In both cases the results 
tend to support a
negative $s$-quark transversity $h_1^s$ within the large uncertainties given.\\
In addition, 
some 
prospects were given for 
measuring precisely the transversity-induced 
polarisation of $\Lambda$ hyperons produced on 
a transversely polarised deuteron target.
Since such a measurement  is 
anticipated to be very sensitive to $h_1^s$, the 
results expected from the upcoming COMPASS run with
a transversely
polarised deuteron target  in the years 2021 and 2022 
will help to improve our knowledge on transversity.

\section*{Acknowledgements}
We gratefully acknowledge the
support of CERN management and staff and the skill and effort of the technicians of our collaborating
institutions.

\newpage

\onecolumn

\section*{Appendix}
Here, the spin transfer for $\Lambda$ and $\bar{\Lambda}$ hyperons is given for the full phase-space and for the current fragmentation region. For each bin the kinematic range is indicated, together with the mean values of $x$, $Q^2$, $z$ and $p_{\rm T}$. 
These and other tables, for all the aforementioned kinematic regions, are available on HEPDATA.\\

\begin{table*}[!h]        
\begin{center}
    \caption{Spin transfer $S_{\Lambda(\bar{\Lambda})}$ from the full phase-space, as a function of $x$, $z$ and $p_{\rm T}$. For each kinematic bin the mean values of $x$, $Q^2$, $z$ and $p_{\rm T}$ are also given. }
        \begin{tabular}{ccccccc}
    \multicolumn{7}{c}{Full phase space} \bigstrut \\
    \hline
      {$x$ range} & $\langle x \rangle$ & $\langle Q^{2} \rangle$ & $\langle z \rangle$ & $\langle p_{\rm T} \rangle$ & $S_{\Lambda}$ & $S_{\bar{\Lambda}}$ \bigstrut \\
      & & (GeV/$c$)$^2$ & & (GeV/$c$)\bigstrut  \\
      \hline
      0.003 - 0.013 & 0.009 & 1.49 & 0.20 & 0.60 & ~~~0.014 $\pm$ 0.106 $\pm$ 0.074 & ~~~0.014 $\pm$ 0.145 $\pm$ 0.107 \bigstrut \\
      0.013 - 0.020 & 0.016 & 2.06 & 0.25 & 0.59 & ~~~0.083 $\pm$ 0.104 $\pm$ 0.078 & ~~~0.061 $\pm$ 0.141 $\pm$ 0.108 \bigstrut \\
      0.020 - 0.032 & 0.025 & 2.75 & 0.28 & 0.57 & -- 0.138 $\pm$ 0.096 $\pm$ 0.077 & ~~~0.125 $\pm$ 0.134 $\pm$ 0.107 \bigstrut \\
      0.032 - 0.060 & 0.044 & 4.30 & 0.31 & 0.55 & -- 0.186 $\pm$ 0.089 $\pm$ 0.076 & ~~~0.017 $\pm$ 0.138 $\pm$ 0.102 \bigstrut  \\
      0.060 - 0.210 & 0.104 & 9.54 & 0.32 & 0.52 & -- 0.101 $\pm$ 0.105 $\pm$ 0.080 & -- 0.122 $\pm$ 0.169 $\pm$ 0.132 \bigstrut \\
      0.210 - 0.700 & 0.290 & 26.5 & 0.34 & 0.53 & ~~~0.074 $\pm$ 0.138 $\pm$ 0.101 & -- 0.399 $\pm$ 0.224 $\pm$ 0.193 \bigstrut \\
      \hline
      {$z$ range} & $\langle x \rangle$ & $\langle Q^{2} \rangle$ & $\langle z \rangle$ & $\langle p_{\rm T} \rangle$ & $S_{\Lambda}$ & $S_{\bar{\Lambda}}$ \bigstrut \\
      & & (GeV/$c$)$^2$ & & (GeV/$c$)\bigstrut \\
      \hline
      0.00 - 0.12 & 0.023 & 4.15 & 0.09 & 0.55 & -- 0.097 $\pm$ 0.154 $\pm$ 0.114 & ~~~0.015 $\pm$ 0.222 $\pm$ 0.163 \bigstrut \\
      0.12 - 0.20 & 0.031 & 4.14 & 0.16 & 0.57 & -- 0.092 $\pm$ 0.095 $\pm$ 0.073 & ~~~0.117 $\pm$ 0.123 $\pm$ 0.099 \bigstrut \\
      0.20 - 0.30 & 0.041 & 4.13 & 0.25 & 0.57 & -- 0.038 $\pm$ 0.083 $\pm$ 0.060 & -- 0.005 $\pm$ 0.113 $\pm$ 0.083 \bigstrut \\
      0.30 - 0.42 & 0.050 & 4.11 & 0.35 & 0.56 & -- 0.152 $\pm$ 0.087 $\pm$ 0.072 & -- 0.205 $\pm$ 0.136 $\pm$ 0.114 \bigstrut \\
      0.42 - 1.00 & 0.058 & 3.99 & 0.53 & 0.58 & ~~~0.118 $\pm$ 0.090 $\pm$ 0.071 & ~~~0.127 $\pm$ 0.173 $\pm$ 0.136 \bigstrut  \\
      \hline 
      {$p_{\rm T} $ range} & $\langle x \rangle$ & $\langle Q^{2} \rangle$ & $\langle z \rangle$ & $\langle p_{\rm T} \rangle$ & $S_{\Lambda}$ & $S_{\bar{\Lambda}}$ \bigstrut \\
      (GeV/$c$) &  & (GeV/$c$)$^2$ & & (GeV/$c$) \bigstrut \\
      \hline
      0.00 - 0.30 & 0.045 & 4.33 & 0.27 & 0.19 & -- 0.079 $\pm$ 0.101 $\pm$ 0.076 & ~~~0.066 $\pm$ 0.158 $\pm$ 0.120 \bigstrut \\
      0.30 - 0.50 & 0.042 & 4.20 & 0.26 & 0.40 & -- 0.101 $\pm$ 0.082 $\pm$ 0.064 & ~~~0.016 $\pm$ 0.114 $\pm$ 0.085 \bigstrut \\
      0.50 - 0.75 & 0.039 & 4.02 & 0.26 & 0.62 & -- 0.066 $\pm$ 0.079 $\pm$ 0.059 & -- 0.002 $\pm$ 0.115 $\pm$ 0.084 \bigstrut \\
      0.75 - 1.10 & 0.036 & 3.91 & 0.27 & 0.89 & ~~~0.068 $\pm$ 0.099 $\pm$ 0.074 & -- 0.054 $\pm$ 0.145 $\pm$ 0.110 \bigstrut \\
      1.10 - 3.50 & 0.034 & 3.97 & 0.28 & 1.35 & -- 0.039 $\pm$ 0.172 $\pm$ 0.122 & -- 0.104 $\pm$ 0.271 $\pm$ 0.206 \bigstrut \\
      \hline 
    \end{tabular}
     \label{tab:table1}
\end{center}{}
\end{table*}
%
\begin{table*}[!h]        
\begin{center}
    \caption{Spin transfer $S_{\Lambda(\bar{\Lambda})}$ from the current fragmentation region ($z \geq 0.2$, $x_F>0$), as a function of $x$, $z$ and $p_{\rm T}$. For each kinematic bin the mean values of $x$, $Q^2$, $z$ and $p_{\rm T}$ are also given. }
        \begin{tabular}{ccccccc}
    \multicolumn{7}{c}{Current fragmentation region} \bigstrut \\
    \hline
      {$x$ range} & $\langle x \rangle$ & $\langle Q^{2} \rangle$ & $\langle z \rangle$ & $\langle p_{\rm T} \rangle$ & $S_{\Lambda}$ & $S_{\bar{\Lambda}}$ \bigstrut \\
      & & (GeV/$c$)$^2$ & & (GeV/$c$) \bigstrut \\
      \hline
      0.003 - 0.013 & 0.009 & 1.42 & 0.31 & 0.62 & ~~~0.024 $\pm$ 0.174 $\pm$ 0.104 & -- 0.190 $\pm$ 0.241 $\pm$ 0.173 \bigstrut \\
      0.013 - 0.020 & 0.016 & 1.81 & 0.33 & 0.60 & ~~~0.212 $\pm$ 0.136 $\pm$ 0.096 & ~~~0.088 $\pm$ 0.184 $\pm$ 0.128 \bigstrut \\
      0.020 - 0.032 & 0.026 & 2.31 & 0.35 & 0.57 & -- 0.110 $\pm$ 0.115 $\pm$ 0.075 & ~~~0.148 $\pm$ 0.164 $\pm$ 0.119 \bigstrut \\
      0.032 - 0.060 & 0.044 & 3.60 & 0.37 & 0.54 & -- 0.169 $\pm$ 0.103 $\pm$ 0.073 & ~~~0.096 $\pm$ 0.169 $\pm$ 0.119 \bigstrut \\
      0.060 - 0.210 & 0.105 & 8.19 & 0.38 & 0.51 & -- 0.110 $\pm$ 0.118 $\pm$ 0.077 & -- 0.303 $\pm$ 0.203 $\pm$ 0.156 \bigstrut \\
      0.210 - 0.700 & 0.290 & 23.4 & 0.38 & 0.53 & ~~~0.122 $\pm$ 0.152 $\pm$ 0.098 & -- 0.448 $\pm$ 0.276 $\pm$ 0.215 \bigstrut \\
      \hline
      {$z$ range} & $\langle x \rangle$ & $\langle Q^{2} \rangle$ & $\langle z \rangle$ & $\langle p_{\rm T}\rangle$ & $S_{\Lambda}$ & $S_{\bar{\Lambda}}$ \bigstrut \\
      & & (GeV/$c$)$^2$ & & (GeV/$c$) \bigstrut \\
      \hline
      0.20 - 0.30 & 0.040 & 4.12 & 0.25 & 0.57 & -- 0.039 $\pm$ 0.083 $\pm$ 0.052 & -- 0.003 $\pm$ 0.113 $\pm$ 0.074 \bigstrut \\
      0.30 - 0.42 & 0.050 & 4.12 & 0.35 & 0.56 & -- 0.152 $\pm$ 0.087 $\pm$ 0.063 & -- 0.202 $\pm$ 0.136 $\pm$ 0.104 \bigstrut \\
      0.42 - 1.00 & 0.058 & 3.99 & 0.53 & 0.58 & ~~~0.119 $\pm$ 0.090 $\pm$ 0.062 & ~~~0.126 $\pm$ 0.173 $\pm$ 0.123 \bigstrut \\
      \hline 
      {$p_{\rm T} $ range} & $\langle x \rangle$ & $\langle Q^{2} \rangle$ & $\langle z \rangle$ & $\langle p_{\rm T} \rangle$ & $S_{\Lambda}$ & $S_{\bar{\Lambda}}$ \bigstrut \\
      (GeV/$c$) & & (GeV/$c$)$^2$ & & (GeV/$c$) \bigstrut \\
      \hline
      0.00 - 0.30 & 0.052 & 4.24 & 0.35 & 0.19 & ~~~0.056 $\pm$ 0.117 $\pm$ 0.073 & -- 0.007 $\pm$ 0.199 $\pm$ 0.131 \bigstrut \\
      0.30 - 0.50 & 0.051 & 4.19 & 0.35 & 0.40 & -- 0.099 $\pm$ 0.104 $\pm$ 0.069 & ~~~0.036 $\pm$ 0.151 $\pm$ 0.102 \bigstrut \\
      0.50 - 0.75 & 0.047 & 4.01 & 0.35 & 0.62 & -- 0.068 $\pm$ 0.102 $\pm$ 0.065 & -- 0.021 $\pm$ 0.163 $\pm$ 0.109 \bigstrut \\
      0.75 - 1.10 & 0.043 & 3.89 & 0.36 & 0.89 & ~~~0.014 $\pm$ 0.128 $\pm$ 0.076 & -- 0.310 $\pm$ 0.191 $\pm$ 0.149 \bigstrut \\
      1.10 - 3.50 & 0.039 & 3.99 & 0.36 & 1.36 & -- 0.076 $\pm$ 0.219 $\pm$ 0.134 & -- 0.086 $\pm$ 0.338 $\pm$ 0.229 \bigstrut \\
      \hline 
    \end{tabular}
     \label{tab:table2}
\end{center}{}
\end{table*}

\newpage



\end{document}